\documentclass[conference, 9pt]{IEEEtran}
\IEEEoverridecommandlockouts
\usepackage{cite}
\usepackage{subcaption}
\usepackage{amsmath,amssymb,amsfonts}
\usepackage{graphicx}
\usepackage{textcomp}
\usepackage{amsmath}
\usepackage{booktabs} 
\usepackage{multirow}
\usepackage{hyperref}
\usepackage{makecell}
\usepackage{makecell}
\usepackage{{adjustbox}}
\usepackage{url}            
\usepackage{booktabs}       
\usepackage{amsfonts}       
\usepackage{nicefrac}       
\usepackage{microtype}      
\usepackage{xcolor}         
\usepackage{algorithm}
\usepackage{algpseudocode}
\usepackage{stfloats}
\usepackage{amsmath}
\usepackage{tikz}

\usepackage{orcidlink}
\usepackage{comment}
\usepackage{amssymb}
\usepackage{braket}
\usepackage{xcolor}
\usepackage[font=footnotesize]{caption} 
\def\BibTeX{{\rm B\kern-.05em{\sc i\kern-.025em b}\kern-.08em
    T\kern-.1667em\lower.7ex\hbox{E}\kern-.125emX}}
\usepackage[compact]{titlesec}

\begin{document}

\title{ \Huge Quantum Portfolio Optimization with Expert Analysis Evaluation
\vspace{-3pt}
}

\author{
\IEEEauthorblockN{Nouhaila Innan\textsuperscript{1,2}, Ayesha Saleem\textsuperscript{1,2}, Alberto Marchisio\textsuperscript{1,2}, and Muhammad Shafique\textsuperscript{1,2}}
\IEEEauthorblockA{
\textsuperscript{1}eBRAIN Lab, Division of Engineering, New York University Abu Dhabi (NYUAD), Abu Dhabi, UAE\\
\textsuperscript{2}Center for Quantum and Topological Systems (CQTS), NYUAD Research Institute, NYUAD, Abu Dhabi, UAE\\
nouhaila.innan@nyu.edu, as17815@nyu.edu, alberto.marchisio@nyu.edu, muhammad.shafique@nyu.edu\\
}
}

\maketitle

\begin{abstract}
Quantum algorithms have gained increasing attention for addressing complex combinatorial problems in finance, notably portfolio optimization. This study systematically benchmarks two prominent variational quantum approaches, Variational Quantum Eigensolver (VQE) and Quantum Approximate Optimization Algorithm (QAOA), under diverse experimental settings, including different asset universes, ansatz architectures, and circuit depths. Although both methods demonstrate effective cost function minimization, the resulting portfolios often violate essential financial criteria, such as adequate diversification and realistic risk exposure. To bridge the gap between computational optimization and practical viability, we introduce an Expert Analysis Evaluation framework in which financial professionals assess the economic soundness and the market feasibility of quantum-optimized portfolios. Our results highlight a critical disparity between algorithmic performance and financial applicability, emphasizing the necessity of incorporating expert judgment into quantum-assisted decision-making pipelines.

\end{abstract}


\section{Introduction}
In today's rapidly evolving financial landscape, the optimization of investment portfolios plays a crucial role in informed decision-making for both institutional and individual investors. Balancing risk and return under various constraints, such as budget, diversification, and investor preferences, requires sophisticated techniques that can navigate high-dimensional, nonlinear, and discrete solution spaces. Traditional methods in quantitative finance have made significant progress in addressing this challenge \cite{gunjan2023brief}. However, the rise of Quantum Computing (QC) presents a new computational paradigm that offers promising solutions to complex financial problems \cite{herman2023quantum,innan2024financial,innan2024financial1,dutta2024qadqn, innan2024lep,alami2024comparative,innan2024qfnn,choudhary2025hqnn}, particularly in portfolio optimization \cite{rebentrost2024quantum,zaman2024poqa}.

According to Markowitz's seminal work on portfolio theory~\cite{markowitz1952portfolio}, portfolio construction involves selecting an asset allocation that balances expected return against risk, subject to constraints such as risk tolerance and budget. Combinatorial portfolio selection, a core part of this stage, is an NP-hard binary optimization problem~\cite{chen2025ben} where one must choose $k$ assets from a pool of $n$, leading to an exponential growth in the number of potential combinations. This complexity is further compounded by practical constraints, including sector diversification, investor preferences, and legal or ethical exclusions.

While classical solvers can tackle such problems effectively on small scales, they become computationally intensive as the portfolio size increases. QC emerges as a compelling alternative, offering \textit{potentially} superior scalability and performance \cite{zaman2023survey,kashif2024computational}. In particular, the Quantum Approximate Optimization Algorithm (QAOA) and Sampling Variational Quantum Eigensolver (SamplingVQE) have been explored as quantum solutions to Quadratic Unconstrained Binary Optimization (QUBO)-encoded portfolio selection problems~\cite{phillipson2021portfolio}.

However, algorithmic solutions, especially those grounded in quantum optimization, face important limitations. Despite achieving high-quality convergence and favorable performance metrics, such algorithms do not inherently incorporate critical real-world considerations such as economic volatility, geopolitical risks, or investor-specific preferences~\cite{zhou2025quantum}. These factors can drastically alter the viability of a given portfolio, regardless of its algorithmically measured performance. Quantum models typically rely on historical pricing data, which, although informative, cannot fully capture the complexities of evolving market dynamics or align with the subjective expectations of investors.

To address this gap, we propose an Expert Analysis Framework as a complementary post-processing stage. This framework incorporates human financial expertise to review the top-ranked quantum-optimized portfolios. The goal is to ensure that selected portfolios are both statistically sound and economically interpretable, as well as practically feasible and aligned with real-world conditions. By integrating human judgment into the quantum pipeline, we enhance the quality and trustworthiness of the final portfolio decisions beyond what traditional loss functions can deliver alone.
Our novel contribution lies in the integration of quantum optimization algorithms with expert-driven validation, bridging computational efficiency and financial interpretability through:
\begin{itemize}
\item A comparative analysis of different quantum circuit architectures, depths, and qubit counts across two quantum optimization algorithms (QAOA and VQE).
\item Evaluation using real-world financial datasets, relying on actual 2025 historical market data while studying diverse asset selections.
\item Development of an expert-guided evaluation framework to assess the practical validity and interpretability of the resulting quantum-generated portfolios.
\end{itemize}

\section{Background and Related Work}
Portfolio optimization has become a representative use case for evaluating quantum algorithms on real-world combinatorial problems. In particular, its formulation as a Quadratic Unconstrained Binary Optimization (QUBO) problem makes it compatible with quantum annealing and gate-based approaches. Prior studies have explored the performance of quantum algorithms on this problem class, analyzing scalability, solution quality, and robustness using both synthetic and historical financial datasets. Comparative investigations have also examined the effectiveness of heuristic, hybrid, and variational methods in handling domain-specific constraints~\cite{phillipson2021portfolio,hodson2019portfolio,Egger_2020,Milhomem2020analysis}.
Building on these efforts, our work examines how quantum algorithms behave in realistic financial settings. We focus on practical limitations that persist despite algorithmic improvements and introduce a complementary framework that integrates expert evaluation into the quantum decision-making process.

SamplingVQE~\cite{peruzzo2014variational,buonaiuto2023best}.
In the Noisy Intermediate-Scale Quantum (NISQ) era, QC has introduced new algorithmic frameworks for addressing discrete optimization problems. Among the most actively studied are Variational Quantum Algorithms (VQAs), which use hybrid quantum-classical routines to optimize parameterized quantum circuits. Two notable VQAs applicable to binary optimization problems, especially portfolio selection, are the QAOA~\cite{farhi2014quantum} and the SamplingVQE~\cite{peruzzo2014variational,buonaiuto2023best}.

QAOA is specifically designed to solve combinatorial optimization problems by encoding a classical cost function into a parameterized quantum circuit. It alternates between two Hamiltonians: the cost Hamiltonian $H_C$, which encodes the problem objective (e.g., a QUBO formulation), and the mixer Hamiltonian $H_M$, which introduces state transitions. The circuit applies alternating layers of these Hamiltonians, parameterized by angles $\boldsymbol{\gamma}$ and $\boldsymbol{\beta}$, starting from an initial state. These parameters are optimized using classical routines to minimize the expectation value of $H_C$. QAOA has been shown to approximate classical solutions with bounded-depth circuits and is considered well-suited for problems with binary constraints.

SamplingVQE, originally developed for estimating molecular ground-state energies in quantum chemistry, has been adapted to optimization tasks by expressing the objective function as a Hamiltonian and minimizing its expectation value over a parameterized quantum state. The algorithm constructs a variational circuit $U(\boldsymbol{\theta})$ with trainable parameters, acting on an initial state, and evaluates the cost via repeated quantum measurements. Unlike QAOA, SamplingVQE offers greater flexibility in circuit design through custom ansatz and can be more easily integrated with classical preprocessing techniques.

Both QAOA and SamplingVQE have been applied to portfolio optimization due to their compatibility with binary decision variables and constraint encoding. Prior studies have demonstrated their feasibility on simulators and early-stage quantum hardware~\cite{phillipson2021portfolio}, yielding near-optimal solutions under realistic constraints in small problem instances.

\section{Methodology}

\begin{figure}[htbp]
    \centering
    \includegraphics[width=1\linewidth]{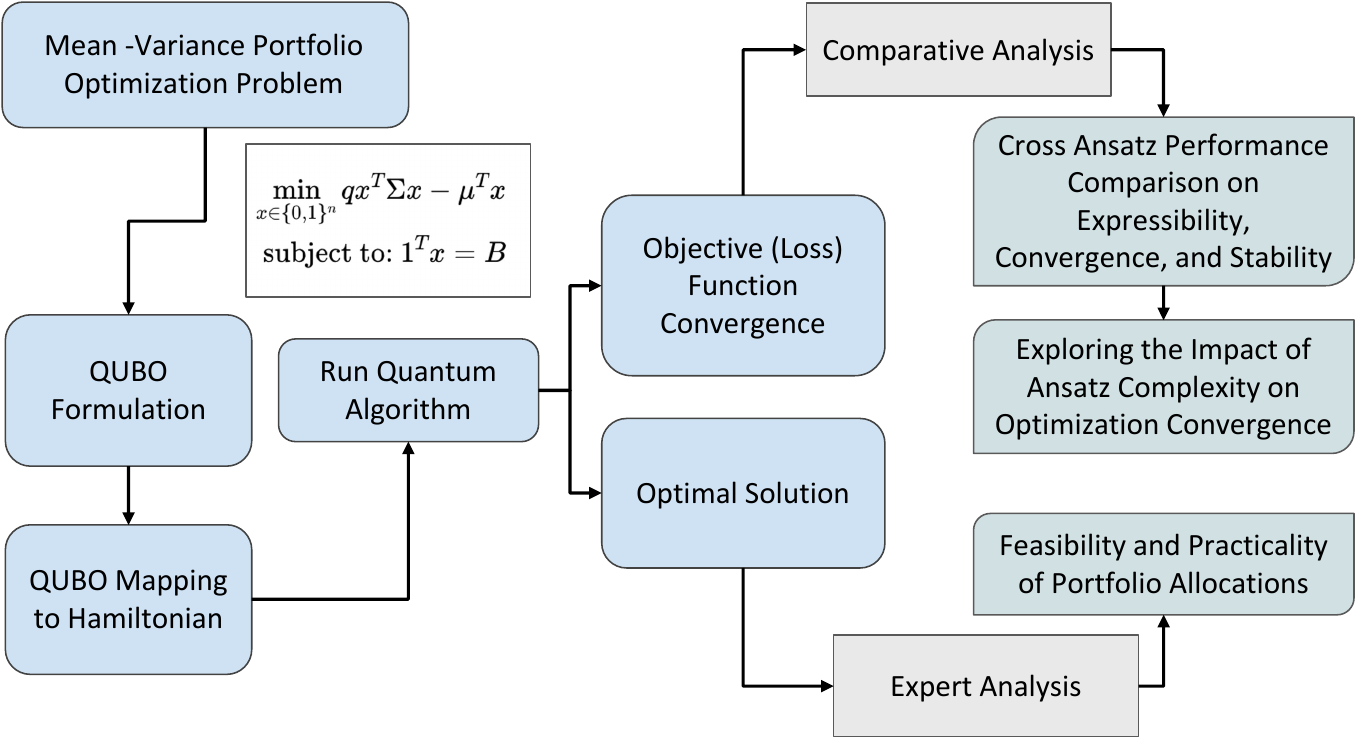}
    \vspace{-0.3cm}
    \caption{Overview of the proposed methodology combining QUBO-based quantum optimization with expert-guided evaluation.}
    \label{method}
\end{figure}

The methodological pipeline begins with the mathematical formulation of the portfolio selection problem as a QUBO instance, proceeds through quantum optimization using variational algorithms, and concludes with an expert evaluation stage to ensure the financial soundness of the resulting portfolios. The entire process is represented in Fig.~\ref{method}.

\subsection{QUBO-based Portfolio Optimization}

The portfolio optimization problem can be formulated as a binary quadratic model. The objective is to select $B$ assets out of $n$ available options to minimize risk and maximize expected return. This leads to the following constrained problem:
$\min_{x \in \{0,1\}^n} \quad q\, x^\top \Sigma x - \mu^\top x$
subject to $ \mathbf{1}^\top x = B,$
where $x \in \{0,1\}^n$ denotes the binary selection vector (i.e., $x_i = 1$ indicates selection of asset $i$), $\mu \in \mathbb{R}^n$ is the vector of expected returns, $\Sigma \in \mathbb{R}^{n \times n}$ is the covariance matrix of asset returns, $q > 0$ is the risk-aversion parameter, and $B$ is the total number of assets to be selected.

Assuming all assets have equal unit price and the full budget must be used, the constraint $\mathbf{1}^\top x = B$ can be incorporated into the objective via a penalty term:
$\min_{x \in \{0,1\}^n} \quad q\, x^\top \Sigma x - \mu^\top x + \alpha \left( \mathbf{1}^\top x - B \right)^2,$
where $\alpha > 0$ controls the strength of the penalty. Expanding the penalty term:
$\left( \sum_{i=1}^n x_i - B \right)^2 = \sum_{i=1}^n x_i + 2 \sum_{i < j} x_i x_j - 2B \sum_{i=1}^n x_i + B^2.$

Using the identity $x_i^2 = x_i$ for binary variables, the full objective becomes a QUBO problem of the form:
$\min_{x \in \{0,1\}^n} \quad x^\top Q x + c,$
where $Q$ is the QUBO matrix with entries:
\[
Q_{ij} =
\begin{cases}
q\, \Sigma_{ij} + \alpha, & \text{if } i \ne j, \\
q\, \Sigma_{ii} - \mu_i + \alpha(1 - 2B), & \text{if } i = j,
\end{cases}
\]
and $c = \alpha B^2$ is a constant offset that can be ignored during optimization.

\subsection{Quantum Algorithm Implementation}
To solve the formulated QUBO problem, we employ QAOA and SamplingVQE. In both approaches, the QUBO objective function $x^\top Q x$ is mapped to a cost Hamiltonian $H_C$ acting on $n$ qubits. This is done using the standard transformation $x_i = \frac{1 - Z_i}{2}$, where $Z_i$ is the Pauli-$Z$ operator on qubit $i$.

Applying this substitution, the cost Hamiltonian takes the form: $H_C = \sum_i h_i Z_i + \sum_{i < j} J_{ij} Z_i Z_j + \text{const},$ where the coefficients $h_i$ and $J_{ij}$ are derived from the QUBO matrix $Q$ by applying the mapping to each quadratic term $Q_{ij} x_i x_j$.

QAOA prepares a parameterized quantum state by alternating between evolution under the cost Hamiltonian $H_C$ and a mixing Hamiltonian $H_M = \sum_i X_i$, where $X_i$ is the Pauli-$X$ operator. A $p$-layer QAOA circuit generates the state:
$|\psi(\boldsymbol{\gamma}, \boldsymbol{\beta})\rangle = \prod_{l=1}^{p} e^{-i \beta_l H_M} e^{-i \gamma_l H_C} |\psi_0\rangle,$
where $|\psi_0\rangle = \frac{1}{\sqrt{2^n}} \sum_{z \in \{0,1\}^n} |z\rangle$ is the uniform superposition over all bitstrings. The variational parameters $\boldsymbol{\gamma}, \boldsymbol{\beta} \in \mathbb{R}^p$ are optimized to minimize the expectation value:
$C(\boldsymbol{\gamma}, \boldsymbol{\beta}) = \langle \psi(\boldsymbol{\gamma}, \boldsymbol{\beta}) | H_C | \psi(\boldsymbol{\gamma}, \boldsymbol{\beta}) \rangle.$

SamplingVQE uses a general parameterized circuit $U(\boldsymbol{\theta})$ to prepare a trial state:
$|\psi(\boldsymbol{\theta})\rangle = U(\boldsymbol{\theta}) |0\rangle^{\otimes n},$
and estimates the objective function by sampling measurement outcomes:
$C(\boldsymbol{\theta}) = \sum_{z \in \{0,1\}^n} P_{\boldsymbol{\theta}}(z) \cdot \text{Cost}(z),$
where $P_{\boldsymbol{\theta}}(z)$ is the probability of observing bitstring $z$, with $z \in \{0,1\}^n$ representing a candidate solution $x$, and $\text{Cost}(z) = z^\top Q z$ is the QUBO cost evaluated classically. The circuit parameters $\boldsymbol{\theta}$ are updated using a classical optimizer to minimize $C(\boldsymbol{\theta})$.

For SamplingVQE, we evaluate several ansatz circuits provided in Qiskit: \texttt{TwoLocal} (with \texttt{Rx}, \texttt{Ry}, and \texttt{Cz} gates), \texttt{EfficientSU2}, \texttt{PauliTwoDesign}, and \texttt{RealAmplitudes}. These differ in circuit depth, entanglement structure, and gate composition, which affects their expressibility and optimization performance~\cite{buonaiuto2023best}.

\subsection{Optimization Evaluation and Convergence Analysis}
To evaluate optimization performance, each ansatz is tested across multiple random seeds and circuit depths. This variation captures the effects of stochasticity in parameter initialization. Convergence behavior is assessed using the cost difference between successive iterations:
$\Delta C_t = C(\boldsymbol{\theta}_{t}) - C(\boldsymbol{\theta}_{t-1}),$
which serves as an indicator of optimization stability and convergence rate. The evolution of the cost function $C(\boldsymbol{\theta}_t)$ over iterations is used to characterize the optimization landscape for each ansatz.

\subsection{Expert Analysis Framework}

Although quantum algorithms produce portfolios optimized under historical data, they do not account for evolving real-world conditions such as geopolitical instability or market-specific risks. To assess the practical viability of the solutions, we incorporate an expert evaluation phase.
In our methodology, domain experts assess the top-ranked quantum-generated portfolios for feasibility in terms of liquidity, sector diversification, and investor suitability. This ensures that the final selection not only satisfies mathematical criteria but also aligns with current market dynamics and investment policy constraints. In this way, expert insight complements quantum outputs with contextual financial reasoning.

First, we analyze algorithmic portfolio return by calculating stock price return of each asset $i$ using $R_i = \left( \frac{P_{i,\text{end}} - P_{i,\text{start}}}{P_{i,\text{start}}} \right) \times 100,$
where ${P_{i,\text{end}}}$ is the stock price of the asset at the end of the time period and ${P_{i,\text{start}}}$ is the stock price of the asset at the beginning of the time period, and then we calculate the average stock return of the portfolio of $n$ assets using the following:
$\text{Average Return (\%)} = \frac{1}{n} \sum_{i=1}^{n} R_i.$
Finally, we evaluate the market feasibility of algorithm-generated portfolios by testing them on stock returns from a ``future'' time period that follows the original set of months used by the system.
\section{Results and Discussion}

\subsection{Experimental Setup}

\begin{figure}[htbp]
    \centering
    \includegraphics[width=1.\linewidth]{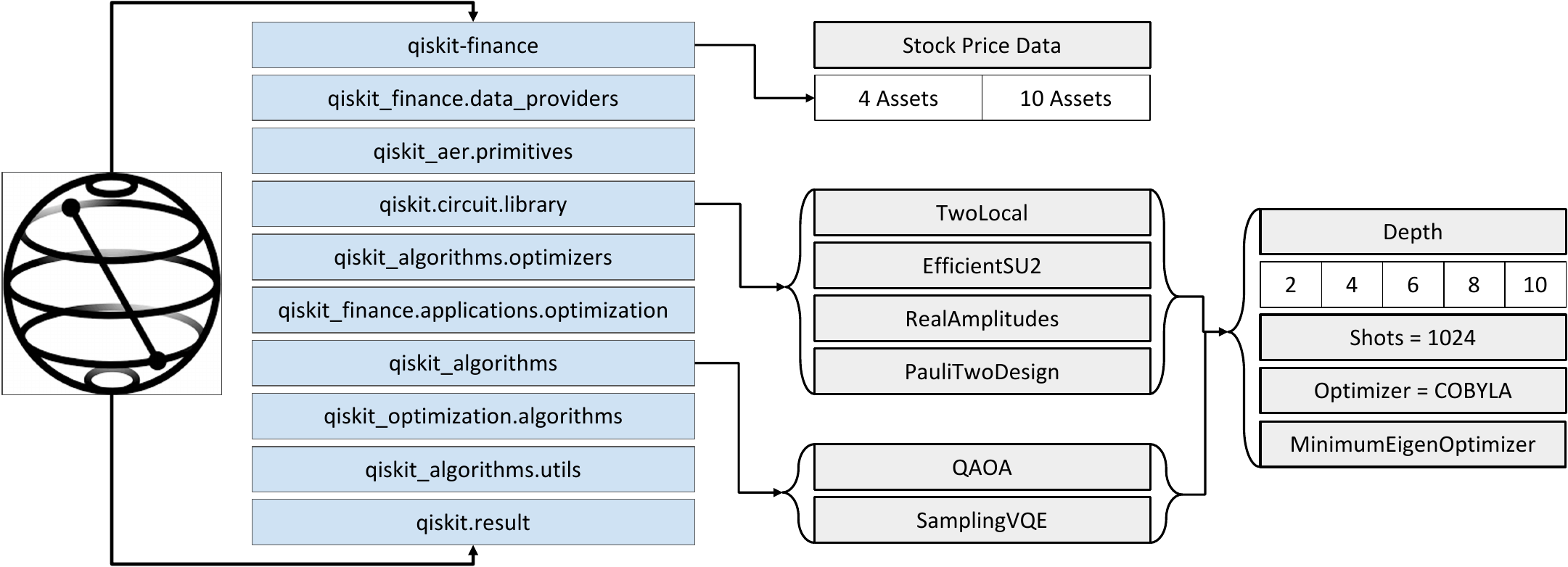}  
    \caption{Experimental setup of our methodology.}
    \label{exp}
\end{figure}

As presented in Fig. \ref{exp}, our experiments are conducted on two problem sizes involving 4 and 10 assets, corresponding to quantum circuits with 4 and 10 qubits, respectively. Historical financial data is collected from Yahoo Finance \cite{yfinance}, covering the six-month period from December 2024 to May 2025. The selected assets are as follows: for the 4-asset portfolio, Apple (AAPL), Google (GOOG), Microsoft (MSFT), and Tesla (TSLA); and for the 10-asset portfolio, the same four assets plus Amazon (AMZN), NVIDIA (NVDA), Goldman Sachs (GS), Morgan Stanley (MS), Nike (NKE), and Coca Cola (KO).

QAOA is implemented using its built-in alternating operator ansatz. SamplingVQE is evaluated using four ansatz architectures: \texttt{TwoLocal} (configured with \texttt{Rx}, \texttt{Ry}, and \texttt{Cz} gates), \texttt{EfficientSU2}, \texttt{PauliTwoDesign}, and \texttt{RealAmplitudes}. All circuits are configured with full entanglement to increase expressibility.
To analyze the impact of circuit complexity, each configuration is executed across five circuit depths: 2, 4, 6, 8, and 10 layers. Every experiment is repeated using five different random seeds. The number of measurement shots is fixed at 1024, and all simulations are performed using the QASM simulator \cite{javadi2024quantum}.

The QUBO penalty parameter $\alpha$ is set to $n/2$ for each instance, where $n$ is the number of assets (and thus qubits). The risk-aversion parameter is fixed at $q = 0.5$, corresponding to a moderate investor profile balancing risk and return.
\subsection{Asset Allocation and Diversification Trends}
Before analyzing the portfolios generated by quantum algorithms, it is important to evaluate the underlying assets in terms of recent performance and their role in diversification. Table~\ref{table1} summarizes the 6-month returns for each asset from December 2024 to May 2025.

Among the 4-asset configuration, Microsoft achieves the highest return (7.23\%), while Apple experienced the steepest decline (-15.97\%). Tesla also posted a loss and is typically associated with high volatility. Google showed a small gain (0.15\%) with relatively low risk. These characteristics suggest that Microsoft and Google are better suited for inclusion in risk-neutral portfolios.

In the broader 10-asset configuration, Coca Cola outperforms all other assets with a 14.11\% gain, providing valuable exposure to the consumer sector. Goldman Sachs and Morgan Stanley show stable returns, making them attractive for financial diversification. Although Amazon and NVIDIA had slight losses, their inclusion can help spread risk across sectors.

Portfolio construction requires balancing two key factors: return potential and diversification. A well-diversified portfolio reduces correlation risk, preventing concentrated losses in a single sector. Assets with stable or modest positive returns are especially appealing for risk-neutral investors, as they imply lower volatility and more consistent future performance.
\begin{table}[ht]
\centering

\caption{6-month return for each asset.}
\begin{adjustbox}{max width=\linewidth}
\begin{tabular}{lrrr}
\toprule
\textbf{Ticker} & \makecell{\textbf{Stock Price} \\ \textbf{(2 Dec 2024)}} & \makecell{\textbf{Stock Price} \\ \textbf{(30 May 2025)}} & \makecell{\textbf{6-Month Return} \\ \textbf{ (\%)}} \\
\midrule
AAPL & 239.013428 & 200.850006 & -15.9670619 \\
GOOG & 172.380157 & 172.642487 & 0.152181089 \\
MSFT & 429.329376 & 460.359985 & 7.22769294 \\
TSLA & 357.089996 & 346.459991 & -2.976842006 \\
AMZN & 210.710007 & 205.009995 & -2.705145371 \\
NVDA & 138.598068 & 135.120621 & -2.509015494\\
GS   & 595.771423 & 600.450012 & 0.785299331 \\
MS   & 129.127823 & 128.029999 & -0.85018393 \\
NKE  & 78.172211  & 60.189999  & -23.00333043 \\
KO   & 62.737671  & 71.590988  & 14.11164434 \\
\bottomrule
\end{tabular}
\end{adjustbox}
\label{table1}
\end{table}
\subsection{Convergence Analysis by Ansatz and Depth}

By varying the ansatz architecture and increasing circuit depth, we assess convergence efficiency and stability under different problem sizes.

\subsubsection{4-Asset Configuration}
\begin{figure*}[htpb]
    \centering
    \includegraphics[width=\linewidth]{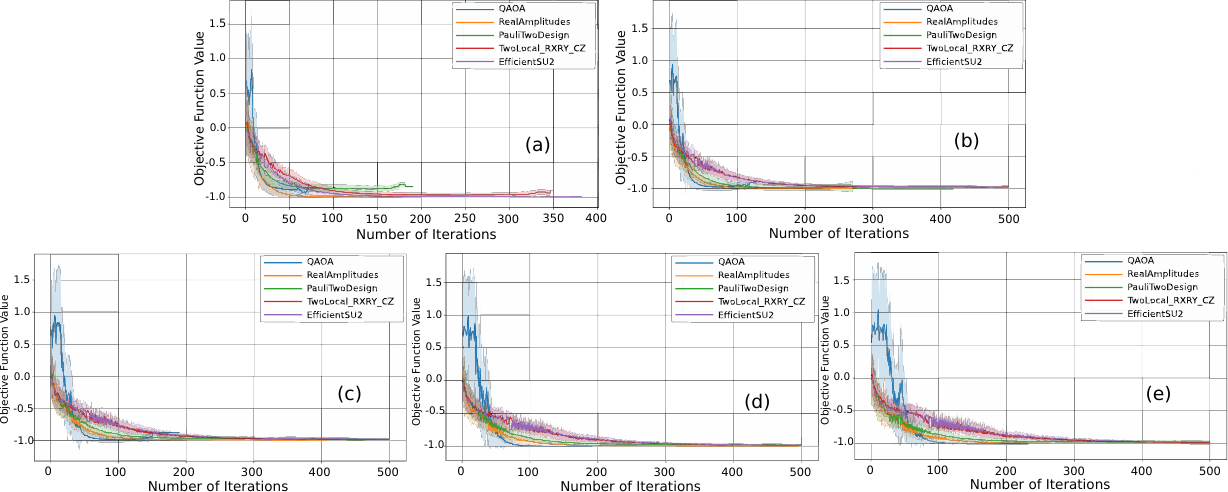}
    \caption{Loss convergence curves for a 4-asset configuration across different circuit depths. (a) Circuit Depth = 2, (b) Circuit Depth = 4, (c) Circuit Depth = 6, (d) Circuit Depth = 8, (e) Circuit Depth = 10.}
    \label{4assetsloss}
\end{figure*}

For the 4-asset configuration, as shown in Fig.~\ref{4assetsloss}, the loss exhibits an overall downward trend across all ansatz architectures. However, a closer analysis reveals notable fluctuations at different circuit depths.
Among all architectures, QAOA exhibits the most instability, with pronounced variance across depths and inconsistent convergence behavior. In contrast, RealAmplitudes, TwoLocal, EfficientSU2, and PauliTwoDesign exhibit smoother and more stable convergence, with only occasional minor fluctuations that do not significantly impact the overall downward trend.

Notably, PauliTwoDesign shows clear improvement with increasing circuit depth, converging from approximately $-0.85$ at depth $=2$ to around $-1.0$ at depth $=10$. RealAmplitudes emerges as the most efficient, consistently reaching convergence in fewer than 100 evaluations across all depths.
Additionally, with increasing depth, all ansatz architectures require more evaluations to converge. This suggests that early convergence in shallower circuits may reflect premature minimization due to insufficient expressibility, rather than true optimization.
\subsubsection{10-Asset Configuration}

\begin{figure*}[htbp]
    \centering
    \includegraphics[width=\linewidth]{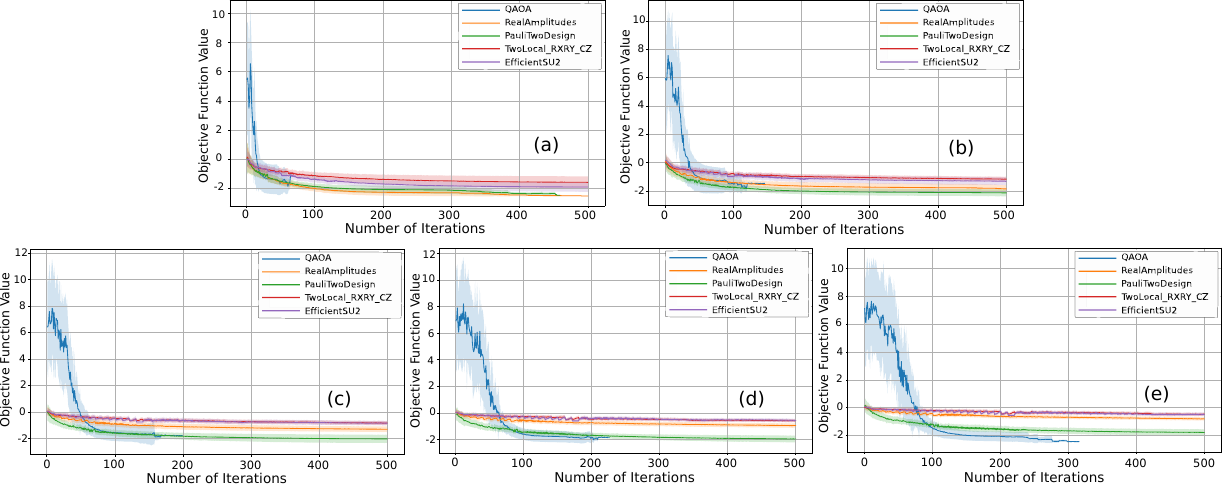}
    \caption{Loss convergence curves for a 10-asset configuration across different circuit depths. (a) Circuit Depth = 2, (b) Circuit Depth = 4, (c) Circuit Depth = 6, (d) Circuit Depth = 8, (e) Circuit Depth = 10.}
    \label{fig4}
\end{figure*}

As shown in Fig.~\ref{fig4}, QAOA continues to exhibit high variance in the 10-asset setup, particularly at lower depths. However, its performance gradually improves with increased depth, evolving from values near $-2.0$ to more stable convergence below $-2.0$ at depth $=10$.
Other ansatz architectures perform more consistently. PauliTwoDesign maintains convergence near $-2.0$ across all depths, while RealAmplitudes, TwoLocal, and EfficientSU2 converge more modestly, typically between $-0.5$ and $-1.0$.

Interestingly, the variance of convergence changes noticeably with depth. For TwoLocal, PauliTwoDesign, and RealAmplitudes, the spread of objective values narrows at greater depths, indicating more stable convergence. In contrast, QAOA shows increasing variance, suggesting sensitivity to circuit depth and parameter initialization.
\subsection{Portfolio Output Analysis}

While convergence results provide insight into an algorithm’s optimization behavior, they do not guarantee the generation of financially viable portfolios. To address this limitation, we analyze the most frequently sampled bitstrings, each representing a candidate portfolio, using probability histograms. These visualizations reveal which portfolios each ansatz tends to favor and how likely they are to be observed under repeated sampling.

\subsubsection{4-Asset Configuration}
This configuration includes firms that, while covering different sub-sectors, are primarily technology-oriented, introducing considerable correlation risk. As a result, optimal portfolios in this setting must strike a balance between return potential and volatility.
\begin{figure*}[htpb]
    \centering
\includegraphics[width=\linewidth]{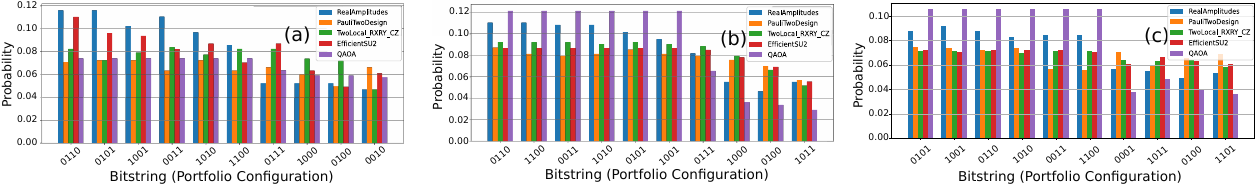}
    \caption{Probability histograms for a 4-asset configuration across different circuit depths. (a) Circuit Depth = 2, (b) Circuit Depth = 6, (c) Circuit Depth = 10.}    
    \label{fig5}
\end{figure*}
\begin{table}[htpb]
\centering
\caption{Assets selected for 4 assets configuration.}
\begin{adjustbox}{max width=\linewidth}
\begin{tabular}{lcc}
\toprule
\textbf{Ansatz/Model} & \textbf{Portfolio} (\textbf{Depth = 2}) & \textbf{Portfolio} (\textbf{Depth = 10})\\
\midrule
Real Amplitudes & \makecell{ $[$GOOG, MSFT$]$ \\ $[$GOOG, TSLA$]$} & [AAPL, TSLA] \\
PauliTwoDesign & \makecell{Fluctuating, Multiple bitstrings \\ at the same probability} & [GOOG, TSLA] \\
TwoLocal & [MSFT, TSLA] & \makecell{Fluctuating, Multiple bitstrings \\ at the same probability} \\
EfficientSU2 & [GOOG, MSFT] & \makecell{Fluctuating, Multiple bitstrings \\ at the same probability} \\
QAOA & \makecell{Fluctuating, Multiple bitstrings \\ at the same probability} & \makecell{Fluctuating, Multiple bitstrings \\ at the same probability} \\
\bottomrule
\end{tabular}
\end{adjustbox}
\label{tab:bitstring4_portfolios}
\end{table}
As shown in Fig.~\ref{fig5} and Table~\ref{tab:bitstring4_portfolios}, many anstaz architectures produce multiple high-probability bitstrings, reflecting selection uncertainty driven by strong asset correlations. At circuit depth 2, both Real Amplitudes and EfficientSU2 identify the [GOOG, MSFT] portfolio, arguably the most balanced and financially sound pair based on historical returns. However, this pattern does not persist at greater depths. Notably, at depth 10, Real Amplitudes selects a risk-heavy portfolio including Apple and Tesla, despite their underperformance in the observed period.

These results suggest that increasing circuit complexity does not necessarily yield better portfolio quality in small, highly correlated asset sets. Instead, deeper circuits may introduce instability or reduce interpretability.

\subsubsection{10-Asset Configuration}
The 10-asset setup across a broader range of industries offers greater opportunities for effective diversification. In this context, an ideal portfolio balances sectoral exposure while maximizing risk-adjusted returns.
\begin{figure*}[htpb]
    \centering
\includegraphics[width=\linewidth]{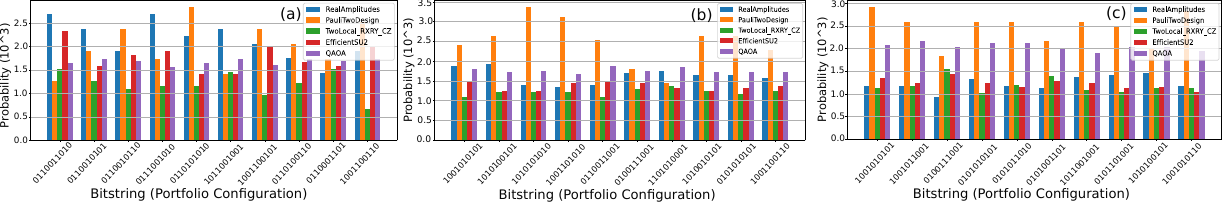}
    \caption{Probability histograms for a 10-asset configuration across different circuit depths. (a) Circuit Depth = 2, (b) Circuit Depth = 6, (c) Circuit Depth = 10.}
\label{fig6}
\end{figure*}
\begin{table}[htpb]
\centering
\caption{Assets selected for 10 assets configuration.}
\begin{adjustbox}{max width=\linewidth}
\begin{tabular}{lcc}
\toprule
\textbf{Ansatz/Model} & \makecell{\textbf{Portfolio} (\textbf{Depth = 2})} & \makecell{\textbf{Portfolio}  (\textbf{Depth = 10})} \\
\midrule
Real Amplitudes & [GOOG, MSFT, NVDA, GS, NKE] & [AAPL, MSFT, AMZN, MS, KO] \\
PauliTwoDesign & [GOOG, MSFT, AMZN, GS, NKE] & [AAPL, TSLA, NVDA, MS, KO] \\
TwoLocal & [GOOG, MSFT, NVDA, GS, NKE] & [GOOG, AMZN, NVDA, GS, KO] \\
EfficientSU2 & [GOOG, MSFT, NVDA, GS, NKE] & [GOOG, AMZN, NVDA, GS, KO] \\
QAOA & \makecell{Fluctuating, Multiple bitstrings \\ at the same probability} & [AAPL, TSLA, NVDA, GS, KO] \\
\bottomrule
\end{tabular}
\end{adjustbox}
\label{tab:ansatz_portfolios}
\end{table}
As illustrated in Fig.~\ref{fig6} and Table~\ref{tab:ansatz_portfolios}, several depth-2 portfolios include Nike, which posts the steepest decline among all assets. In contrast, portfolios generated at depth 10 tend to align more closely with sound financial principles. Coca Cola consistently appears across different ansatz architectures and is often selected alongside Amazon, NVIDIA, or Google to enhance diversification and reduce volatility.

Despite this trend, some architectures, such as Real Amplitudes, PauliTwoDesign, and QAOA, continue to select Apple or Tesla at depth 10, even though both assets experience significant losses. These choices potentially weaken portfolio quality and underscore the risk of relying solely on convergence metrics.
The most balanced outputs are obtained from TwoLocal and EfficientSU2 at depth 10. These circuits consistently generate diversified portfolios that include Google, Coca Cola, Goldman Sachs, and either Amazon or NVIDIA, combining strong sectoral coverage with reasonable return expectations.

These findings emphasize that deeper circuits tend to produce more viable portfolios in larger, more diverse asset sets. However, good convergence alone does not ensure financial suitability. These results highlight the need for interpretability and domain expertise when applying quantum algorithms to portfolio optimization.

\subsection{Expert Evaluation of Portfolio Quality}

To assess the real-world viability of the quantum-generated portfolios, we compare their composition and short-term future returns using our expert analysis. Table~\ref{tab:june2025_returns} reports the individual asset returns during June 2025, the month immediately following the original six-month evaluation period, while Table~\ref{tab:future_portfolio_returns} summarizes the average returns for each generated portfolio during that period, calculated using our framework.

\begin{table}[htpb]
\centering
\caption{Stock returns for each asset in June 2025 (2 June to 20 June), following the 6-month evaluation period.}
\begin{adjustbox}{max width=\linewidth}
\begin{tabular}{lrrr}
\toprule
\textbf{Ticker} & \textbf{Price (2 Jun 2025)} & \textbf{Price (20 Jun 2025)} & \textbf{June Return (\%)} \\
\midrule
AAPL & 201.70 & 201.00 & -0.35 \\
GOOG & 170.17 & 167.73 & -1.43 \\
MSFT & 461.97 & 477.40 & 3.34 \\
TSLA & 342.69 & 322.16 & -5.99 \\
AMZN & 206.65 & 209.69 & 1.47 \\
NVDA & 137.37 & 143.85 & 4.72 \\
GS & 598.72 & 640.80 & 7.03 \\
MS & 128.40 & 132.71 & 3.36 \\
NKE & 61.57 & 59.79 & -2.89 \\
KO  & 71.49 & 68.84 & -3.71 \\
\bottomrule
\end{tabular}
\end{adjustbox}
\label{tab:june2025_returns}
\end{table}

\begin{table}[htpb]
\centering
\caption{Average returns for selected portfolios in June 2025, based on configurations generated during the 6-month period.}
\begin{adjustbox}{max width=\linewidth}
\begin{tabular}{cc|cc}
\multicolumn{2}{c|}{\textbf{4-Asset}} & \multicolumn{2}{c}{\textbf{10-Asset}} \\
\toprule
\textbf{Portfolio} & \textbf{Return (\%)} & \textbf{Portfolio} & \textbf{Return (\%)} \\
\midrule
GOOG, MSFT & 0.95 & GOOG, MSFT, KO, GS, AMZN & 1.34 \\
AAPL, TSLA & -3.17 & GOOG, MSFT, KO, GS, NVDA & 1.99 \\
GOOG, TSLA & -3.71 & GOOG, AMZN, NVDA, GS, KO & 1.62 \\
MSFT, TSLA & -1.33 & AAPL, TSLA, NVDA, MS, KO & -0.39 \\
& & AAPL, MSFT, AMZN, GS, KO & 0.34 \\
\bottomrule
\end{tabular}
\end{adjustbox}
\label{tab:future_portfolio_returns}
\end{table}

In the 4-asset configuration, the only portfolio to generate a positive return in June is [GOOG, MSFT], identified by Real Amplitudes and EfficientSU2 at depth 2. All other combinations, particularly those involving Tesla or Apple, underperform, an outcome that aligns with prior expert observations about volatility and return consistency.

In contrast, the 10-asset portfolios show stronger overall performance, with four out of five configurations producing positive returns. Notably, the highest-returning portfolio, [GOOG, MSFT, KO, GS, NVDA], at 1.99\%, is not generated by any circuit. However, EfficientSU2 generates a similar composition with a 1.62\% return, indicating promising alignment with expert expectations.

From an expert perspective, this analysis supports three key insights:
\begin{itemize}
\item Portfolios containing Microsoft, Google, and Goldman Sachs consistently perform well, reinforcing their classification as stable, lower-risk assets.
\item Deeper circuits (e.g., depth 10) in larger asset configurations are more likely to yield viable portfolios, as they support better diversification.
\item Portfolios containing Tesla or Apple frequently underperform, validating expert caution regarding volatility and recent performance trends.
\end{itemize}
While quantum circuits show clear promise in rapidly generating competitive candidate portfolios, their inherently probabilistic nature sometimes leads to inconsistencies with forward-looking financial expectations. The ability to narrow the solution space is valuable, but expert oversight remains essential for final portfolio selection. This evaluation demonstrates the role of expert analysis as a critical filter, one that ensures practical portfolio quality beyond convergence or historical return metrics.

\subsection{Discussion}
This study addresses the portfolio optimization problem using real financial data over a six-month period, evaluated across two asset configurations using various architectures and depths. While the results demonstrate clear algorithmic convergence, particularly as circuit depth increases, this convergence does not consistently translate into the generation of optimal portfolios.

Although the ansatz architectures exhibit strong scalability and optimization behavior, the portfolios they produce often fail to align with favorable future outcomes. This limitation highlights a critical characteristic of current quantum approaches: they operate on historical stock data alone and do not incorporate forward-looking or external variables that influence asset performance. Moreover, while simulator-based results demonstrate idealized performance, real quantum hardware introduces additional complexity, such as noise and decoherence, which may further affect output quality.

Despite this, QC provides a potential advantage by addressing the NP-hard nature of the portfolio optimization problem. It enables the efficient and scalable generation of candidate portfolios that would otherwise be computationally intensive to explore classically. Building upon this strength, our Expert Analysis framework serves as a complementary layer, allowing further refinement and selection of portfolios based on future expectations, real-time dynamics, and investor-specific constraints.

The results show that quantum-generated portfolios benefit from expert interpretation. Our framework offers a practical path forward, in which quantum algorithms narrow the solution space while domain knowledge ensures financial relevance. It also leaves room for future enhancements that incorporate probabilistic forecasting, alternative risk models, and behavioral or ethical investment preferences.

\section{Conclusion}

This work presents a hybrid framework for quantum-enhanced portfolio optimization, combining quantum computational scalability with expert-informed evaluation. Our results demonstrate that quantum circuits, particularly at higher depths, achieve strong convergence across varying asset configurations and ansatz architectures. However, convergence alone does not guarantee portfolio quality, as future performance is shaped by dynamic and unmodeled external factors.

To bridge this gap, we introduce an Expert Analysis layer that evaluates quantum-generated portfolios beyond historical data. This component enhances interpretability and aligns portfolio selection with financial context and investor needs.
Our findings suggest that QC holds strong potential for addressing complex financial optimization tasks. However, to produce robust and viable outcomes, quantum solutions must integrate domain expertise. The proposed framework lays the foundation for future research into adaptive, expert-guided quantum financial systems capable of responding to both market fluctuations and individual investor preferences.




\section*{Acknowledgment}
This work was supported in part by the NYUAD Center for Quantum and Topological Systems (CQTS), funded by Tamkeen under the NYUAD Research Institute grant CG008.
\bibliographystyle{IEEEtran}
\footnotesize
\bibliography{refs}







\end{document}